\def\rn{\noindent\parshape 2 0truecm 8.8truecm 0.3truecm 8.5truecm}
\def\nn#1 #2{#1, #2.}				
\def\nnn#1 #2 #3{#1, #2. #3.}			
\def\nnnn#1 #2 #3 #4{#1, #2. #3. #4.}		
\def\nnnnn#1 #2 #3 #4 #5{#1, #2. #3. #4. #5.}	
\def\multiand{, \&\hbox{ }}				

\def\rg#1;#2;#3;#4;#5;#6 {\par\rn#1 #2, {\it #3}, {\bf #4}, #5 (``#6'') \par}
\def\rf#1;#2;#3;#4;#5 {\par\rn#1 #2, {\it #3}, {\bf #4}, #5\par}
\def\rfbook#1;#2;#3;#4;#5 {{\frenchspacing\par\rn#1 #2, {\it #3} (#4: #5)\par}}
\def\rfproc#1;#2;#3;#4;#5;#6 {{\frenchspacing\par\rn#1 #2, in {\it #3}, ed. #4 (#5: #6)\par}}
\def\rfprep#1;#2;#3  {{\par\rn#1 #2, #3\par}}
\def\rfprepp#1;#2;#3 {{\par\rn#1 #2, #3\par}}

\def\second{{\rm s}}

\def\milliK{{\rm mK}}
\def\mK{{\rm \mu K}}

\def\expec#1{\langle#1\rangle}

\def\etal{{\frenchspacing\it et al.}}
\def\ie{{\frenchspacing\it i.e.}}
\def\eg{{\frenchspacing\it e.g.}}

\def\rms{rms}

\def\beq#1{\begin{equation}\label{#1}}
\def\eeq{\end{equation}}
\def\beqa#1{\begin{eqnarray}\label{#1}}
\def\eeqa{\end{eqnarray}}
\def\eq#1{equation~(\ref{#1})}

\def\eqn#1{~(\ref{#1})}

\def\fig#1{Figure~\ref{#1}}
\def\Fig#1{Figure~\ref{#1}}

\def\sec#1{Section~\ref{#1}}

\def\spose#1{\hbox to 0pt{#1\hss}}
\def\simlt{\mathrel{\spose{\lower 3pt\hbox{$\mathchar"218$}}
     \raise 2.0pt\hbox{$\mathchar"13C$}}}
\def\simgt{\mathrel{\spose{\lower 3pt\hbox{$\mathchar"218$}}
     \raise 2.0pt\hbox{$\mathchar"13E$}}}
\def\simpropto{\mathrel{\spose{\lower 3pt\hbox{$\mathchar"218$}}
     \raise 2.0pt\hbox{$\propto$}}}

\def\n{\varepsilon}

\def\ed{\end{document}}

\def\ith{i^{th}}

\def\l{\ell}

\def\b{{\bf b}}
\def\x{{\bf x}}
\def\xt{\tilde{\bf x}}
\def\y{{\bf y}}
\def\yt{\tilde{\bf y}}
\def\z{{\bf z}}
\def\n{{\bf n}}
\def\nt{\tilde{\bf n}}
\def\rh{\widehat{\bf r}}
\def\A{{\bf A}}
\def\At{\tilde{\bf A}}
\def\D{{\bf D}}
\def\I{{\bf I}}
\def\M{{\bf M}}
\def\N{{\bf N}}
\def\NN{{\bf\Sigma}}
\def\S{{\bf S}}

\def\B{{\bf B}}
\def\NT{\tilde{\bf N}}
\def\zero{{\bf 0}}

\font\bfmath=cmmib10

\font\bfmath=cmmib10
\def\err{\hbox{\bfmath\char'042}}       

\def\GUM{$\gamma$UMi}
\def\AL{$\alpha$Leo}

\documentstyle[emulateapj,danonecolfloat,epsf]{article}

\begin{document}
\twocolumn[


\journalid{337}{15 January 1989}
\articleid{11}{14}

\submitted{Revised February 13, 2000.}

\title{COSMIC MICROWAVE BACKGROUND MAPS FROM THE 
HACME EXPERIMENT}

\author{
Max Tegmark
\footnote{Dept. of Physics, Univ. of Pennsylvania, 
Philadelphia, PA 19104;
max@physics.upenn.edu}$^,$\footnote{Institute for Advanced Study, 
Princeton, 
NJ 08540}$^,$\footnote{Hubble Fellow},
Ang\'elica de Oliveira-Costa$^{a,b}$,
John W. Staren\footnote{Physics Dept., University 
of California, 
Santa Barbara, CA 93106, and NSF Center for 
Particle Astrophysics},
Peter R. Meinhold$^d$,
Philip M. Lubin$^d$,\\
Jeffrey D. Childers$^d$,
Newton Figueiredo\footnote{Escola Federal de 
Engenharia de Itajub\'{a}, P.O.
Box 50, 37500-000, Itajub\'{a}, MG, Brazil; 
newton@cpd.efei.br},
Todd Gaier\footnote{JPL, 4800 Oak Grove Dr., 
Pasadena, CA 91109},
Mark A. Lim\footnote{Institute for Space and 
Astronautical Science,
3-1-1 Yoshinodai Sagamihara, Kanagawa 229, 
Japan}, 
Michael D. Seiffert$^d$, 
Thyrso Villela\footnote{INPE, S\~ao  Jos\'e dos  
Campos, SP 12201-970, Brazil} 
and
C. Alexandre Wuensche$^h$.
}

\begin{abstract}
We present Cosmic Microwave Background (CMB) maps 
from the Santa Barbara HACME 
balloon experiment (Staren {\etal} 2000), 
covering about 1150 square degrees split between
two regions in the northern sky, near the stars 
$\gamma$ Ursae Minoris and 
$\alpha$ Leonis,
respectively. 
The FWHM of the beam
is $\sim 0.77^\circ$ in 
three frequency bands centered on 39,  41 and 43 
GHz. 
The results demonstrate that the thoroughly 
interconnected scan strategy
employed allows efficient removal of $1/f$-noise 
and slightly 
variable scan-synchronous offsets.
The maps display no striping, and the noise 
correlations are 
found to be virtually isotropic, decaying on an 
angular scale 
$\sim 1^\circ$.
The noise performance of the experiment resulted 
in an upper limit on CMB anisotropy. However, 
our results demonstrate that atmospheric 
contamination and
other systematics resulting from the circular 
scanning strategy can be
accurately controlled, and bodes well for the 
planned follow-up 
experiments BEAST and ACE, since they show that 
even with the 
overly cautious assumption that 
$1/f$-noise and offsets will be as dominant as 
for HACME, the problems 
they pose can be readily overcome with the 
mapmaking algorithm discussed.
Our prewhitened notch-filter algorithm for 
destriping and 
offset removal should be useful also for other 
balloon- and ground-based 
experiments whose scan strategies involve
substantial interleaving.
\end{abstract}

]


\section{INTRODUCTION}

One of the main goals of the next generation of 
Cosmic Microwave Background 
(CMB) experiments is to produce maps that can 
resolve degree scale features 
corresponding to acoustic 
peaks in the angular power spectrum (see {\eg} 
Bond 1996 and 
Hu {\etal} 1997 for reviews), since this 
potentially allows accurate
determination of cosmological parameters 
(Jungman {\etal} 1996; Bond {\etal} 1997; 
Zaldarriaga {\etal} 1997). 

\begin{figure}[phbt]
\centerline{{\vbox{\epsfysize=9.5cm\epsfbox{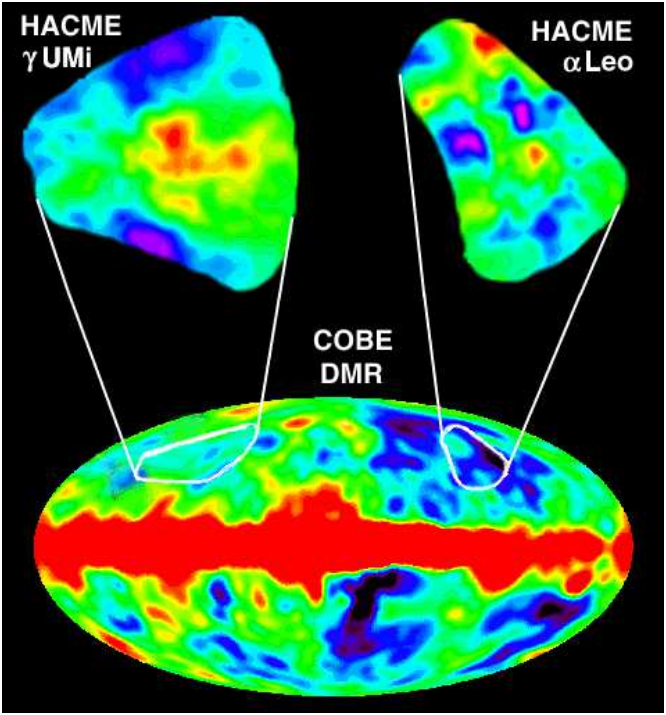}}}}
\caption{
Location of observing regions in galactic 
coordinates.
}
\label{ZoomFig}
\end{figure}

To date, maps with degree scale angular 
resolution 
have been published from
the ACME/MAX (White \& Bunn 1995), 
Saskatoon (Tegmark {\etal} 1997 -- hereafter 
``T97d''), CAT (Jones 1997) 
and QMAP 
(Devlin {\etal} 1998;
Herbig {\etal} 1998;
de Oliveira-Costa {\etal} 1998) and 
Python (Coble {\etal} 1999)
experiments. 
The maps we present are based on the 1996 flight 
of the HACME balloon 
experiment, which is described in detail by 
Staren {\etal} (2000, hereafter ``S2000''). They 
cover regions around the
stars $\gamma$ Ursae Minoris (\GUM) and $\alpha$ 
Leonis (\AL), as
shown in \fig{ZoomFig} (the COBE DMR map is from 
Bennett {\etal} 1996). 
The data analysis we 
describe below is an interesting pathfinder for 
upcoming mapping 
experiments since HACME differed from its 
predecessors 
ACME/MAX, Saskatoon and CAT in 
three 
important ways:
\begin{enumerate}
\item The time-ordered data set is substantially 
larger 
(consisting of $\sim 10^7$ temperature 
measurements).
\item $1/f$-noise plays a significant role.
\item The scan strategy interconnects the 
observed pixels 
in a complicated way.
\end{enumerate}
To use this interconnectedness for efficient 
$1/f$-noise

\smallskip

\noindent
removal, 
we apply the lossless mapmaking 
algorithm described in Tegmark (1997a, hereafter 
``T97a'')
tailored to the noise power
spectrum at hand. Our analysis provides the first 
large scale test of
this method, since the large size of our data set 
forces us to use  
all the numerical tricks that have been proposed 
for accelerating it 
(Wright 1996; Tegmark 1997c -- hereafter 
``T97c''). 
We describe our analysis in Section 2 and present 
the results in Section 3.

\section{METHOD}
\label{MethodSec}

We make our maps using the minimum-variance 
method described by
Wright (1996), T97a and T97c, which we summarize 
briefly below.

Let $y_i$ denote the temperature measured by the 
HACME experiment at the 
$\ith$ observation (there were 312.5 such 
measurements per second), and group
this time-ordered data set (TOD) into an $M$-dimensional 
vector $\y$.
As described in S2000, there are $M=3,323,250$ 
measurements
in the {\GUM} region and 1,743,625 in
the {\AL} region,  respectively, for each of the 
three 
frequency channels.
Discretizing the map into $N$ pixels centered at 
unit vectors
$\rh_1,...,\rh_N$, we let $x_i$ denote the true 
beam-smoothed
CMB temperature in the 
direction $\rh_i$ and group these temperatures 
into an $N$-dimensional
map vector $\x$.
The TOD is related to the map by
\beq{LinearProblemEq}
\y = \A\x+\n,
\eeq
where $\n$ denotes a random noise vector
and the $M\times N$ matrix $\A$ encodes the HACME 
scan strategy.
Letting $N_i$ denote the number of the
pixel pointed to at the $\ith$ observation, 
we have $\A_{ij}=1$ if $N_i=j$, $\A_{ij}=0$ 
otherwise,
so all the entries of $\A$ are zero except that 
there
is a single ``1" in each row.
We find that $\n$ can be accurately modeled as a 
multivariate
Gaussian random variable with  
zero mean ($\expec{\n}=\zero$), so it is 
completely specified by its 
covariance matrix $\N\equiv\expec{\n\n^t}$.

\begin{figure}[phbt]
\centerline{{\vbox{\epsfxsize=9.7cm\epsfbox{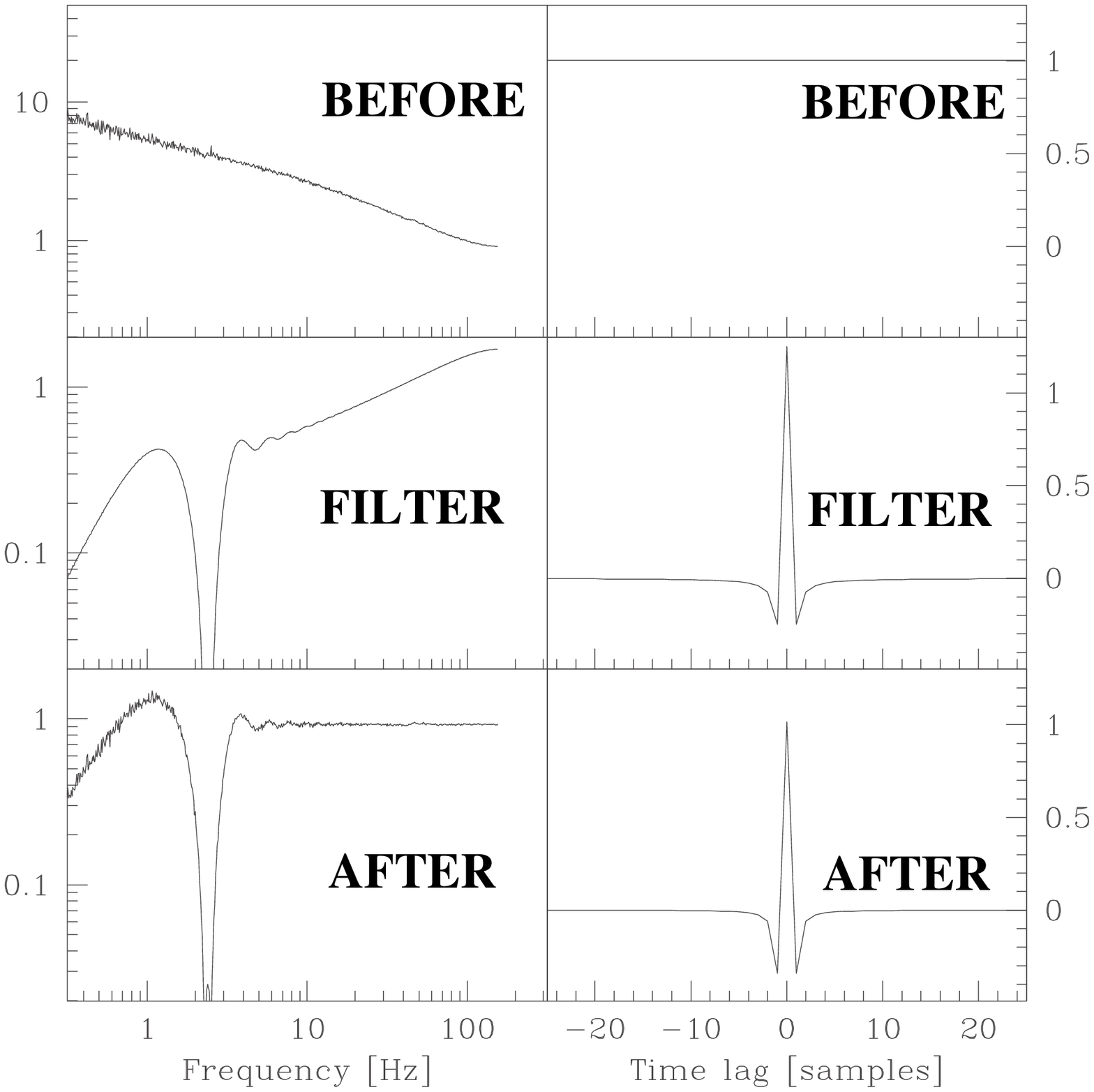}}}}
\caption{
The prewhitening procedure is illustrated in 
the frequency domain (left) and in the time domain
(right) for the {\GUM} 43 GHz channel.
The noise power spectrum (upper left) contains a 
$1/f$-component
which causes the noise to be almost perfectly 
correlated between
measurements close together in time (upper 
right).
By multiplying the Fourier transformed signal 
by an appropriate prewhitening 
filter (middle left), which corresponds to 
applying a 
convolution filter (middle right) to the signal,
the filtered TOD $\yt$ obtains a noise power
spectrum which is close to white noise with notches
at 0 Hz and the spin frequency (lower left), corresponding to a
time autocorrelation function that falls of rapidly (lower right).
}
\label{NoiseFig}
\end{figure}

The problem at hand is to compute a map $\xt$ 
that estimates
the true map $\x$ using the TOD $\y$.
As shown in T97a, all the cosmological information
from the TOD is retained in the map $\xt$ given 
by 
\beq{xtDefEq}
\xt=[\A^t\M\A]^{-1}\A^t\M\y,
\eeq
where the matrix $\M\equiv\N^{-1}$.
Substituting this into \eq{LinearProblemEq} shows 
that we can write 
$\xt=\x+\err$, {\ie}, that our map estimate $\xt$ 
equals the true
map $\x$ plus a Gaussian noise term $\err$ that 
is independent of $\x$.
This holds for any choice of $\M$.
This pixel noise has zero mean 
($\expec{\err}=\zero$) and a covariance 
matrix given by
\beq{SigmaDefEq}
\NN\equiv\expec{\err\err^t} = 
[\A^t\M\A]^{-1}[\A^t\N^{-1}\A][\A^t\M\A]^{-
1}.
\eeq
To minimize the noise, we want to choose 
$\M\approx\N^{-1}$ (at least approximately),
which gives $\NN$ near the best possible case 
$[\A^t\N^{-1}\A]^{-1}$.
For the simple case of white noise from the 
detector, we would have 
$\N\propto\I$, the identity matrix, and $\xt$ 
would be obtained by 
simply averaging the measurements of each pixel 
in the map.
We will refer to this as the {\it straight 
averaging} mapmaking method,
in contrast to the prescription of \eq{xtDefEq}, 
which we will call 
the {\it minimum noise method}.
In our case, the matrix $\N$ is far
from diagonal, since long-term $1/f$ drifts 
introduce correlations between
the noise $n_i$ at different times.
To eliminate the need for a numerically 
unfeasible inversion of $\N$
in equations\eqn{xtDefEq} and\eqn{SigmaDefEq}, we 
employ the prewhitening 
trick of Wright (1996) and define a new data set 
$\yt\equiv\D\y$, where the
prewhitening matrix $\D$ is chosen so that the 
prewhitened noise
$\nt\equiv\D\n$ becomes as close to possible to 
white noise, \ie, 
so that the filtered noise covariance matrix
$\NT\equiv\expec{\nt\nt^t}=\D\N\D^t\approx\I$.
Defining $\At\equiv\D\A$, this allows us to 
rewrite \eq{LinearProblemEq}
as $\yt = \At\x+\nt$, so equations\eqn{xtDefEq} 
and\eqn{SigmaDefEq}
give
\beqa{PrewhitenedxtEq}
\xt&=&[\At^t\M\At]^{-1}\At^t\M\yt\nonumber\\
   &=&[\A^t\D^t\M\D\A]^{-1}\A^t\D^t\M\D\y,\\
\label{PrewhitenedSigmaEq}
\NN&=&[\A^t\D^t\M\D\A]^{-1}
      [\A^t\D^t\M\NT\M^t\D\A]
      [\A^t\D^t\M\D\A]^{-1},\nonumber
\eeqa
where we make the choice $\M\equiv\I$ since 
$\NT\approx\I$.
To accelerate the numerical evaluation of these 
expressions,
we take advantage of the fact that $\D^t\D$ is 
almost
a circulant matrix, as described in detail in 
T97c.
Since the statistical properties of the noise are 
approximately
constant over time 
(the covariance $\expec{n_i n_j}$ depends only on 
the time
interval $|i-j|$ between the two observations), 
the prewhitening 
filter $\D$ is simply a convolution filter. We 
compute it as described
in T97a, and the result is plotted in \fig{NoiseFig}.
The prewhitening procedure  renders the data 
insensitive to the 
mean (monopole) of the map. This makes the matrix 
$[\At^t\M\At]$
non-invertible, which is remedied using the 
pseudo-inverse technique
described in the Appendix of Tegmark (1997b).

\begin{figure*}[phbt]
\centerline{{\vbox{\epsfxsize=18cm\epsfbox{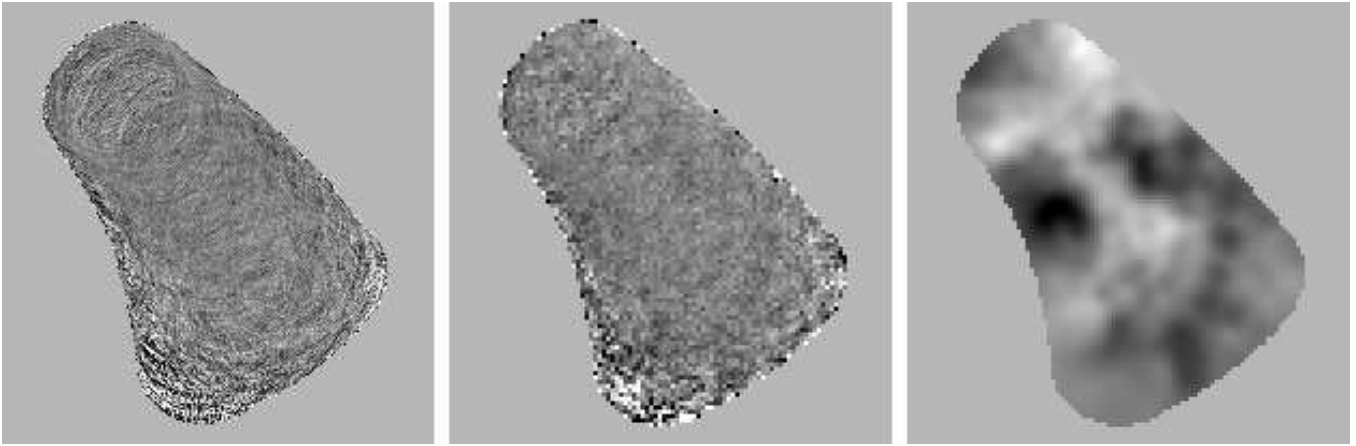}}}}
\caption{
Straight averaging (left), optimal unsmoothed 
(center) and Wiener filtered  
(right) maps of the {\AL} region at 43 GHz.
}
\label{MethodFig}
\end{figure*}

\begin{figure*}[phbt]
\centerline{{\vbox{\epsfxsize=18cm\epsfbox{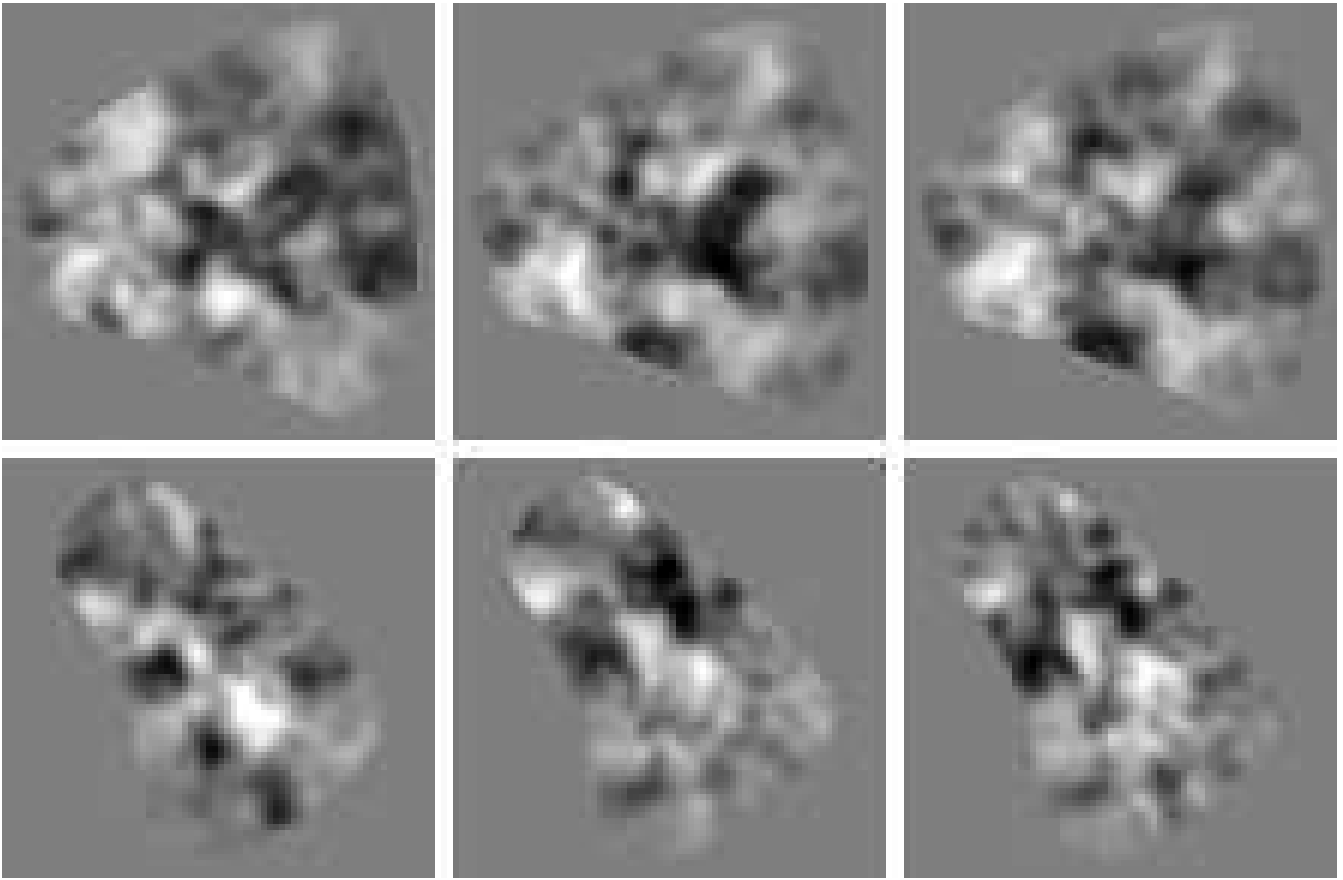}}}}
\caption{
Wiener-filtered maps of the {\GUM} (top) and 
{\AL} (bottom)
regions at 39 GHz (left), 41 GHz (center) and 43 
GHz (right).
}
\label{WienerFig}
\end{figure*}

In addition to the map $\xt$, we also compute the 
{\it Wiener filtered}
map $\x_w$, which is given by (Wiener 1949)
\beq{WienerEq}
\x_w\equiv\S[\S+\N]^{-1}\xt,
\eeq
where the CMB signal covariance matrix 
$\S\equiv\expec{\x\x^t}$   
is given by the standard formula
\beq{SdefEq}
\S_{ij} = \sum_{\l=2}^\infty {(2\l+1)\over 
4\pi}P_\l(\rh_i\cdot\rh_j) B_\l^2 C_\l
\eeq
for some assumed fiducial power spectrum $C_\l$. 
In our analysis, we 
approximate the 
HACME beam as a circular
FWHM$=0.77^\circ$ Gaussian, although the actual 
beam has an asymmetry of order 6\%. 
Thus $B_\l\approx e^{-\theta^2\l(\l+1)/2}$,
where $\theta\equiv (8\ln 2)^{-1/2}$~FWHM.
As discussed in T97a, 
the Wiener filtered map 
$\x_w$ contains the same cosmological information 
as $\xt$ (which we will refer
to as  the {\it unsmoothed} map to distinguish it 
from $\x_w$), but is more 
useful for visual inspection since it is less 
noisy.
Wiener filtering has been successfully applied to 
both CMB maps ({\eg},
Bunn {\etal} 1994, 1996; T97d) and 
galaxy surveys (\eg, 
Fisher {\etal} 1995; Lahav {\etal} 1994; Zaroubi 
{\etal} 1995). 

After removal of calibration data from $\z$, 
segments 
during which control commands were transmitted to 
the gondola
are omitted, which reduces the size of the TOD by 
about $0.4\%$.
The remainder has scan-synchronous offsets 
described in S2000.  
We determine the time-independent scan-
synchronous component 
simultaneously with the map by including 
unknown offsets, one corresponding to each of the
125 sampling positions along the circular scan, 
as 125 additional ``pixels'' in the vector $\x$
and widening $\A$ accordingly with an additional 
``1''
in each row. This trick was subsequently applied
also in the analysis of QMAP 
--- see de Oliveira-Costa {\etal} (1998).

\begin{figure}[phbt]
\centerline{{\vbox{\epsfxsize=8.9cm\epsfbox{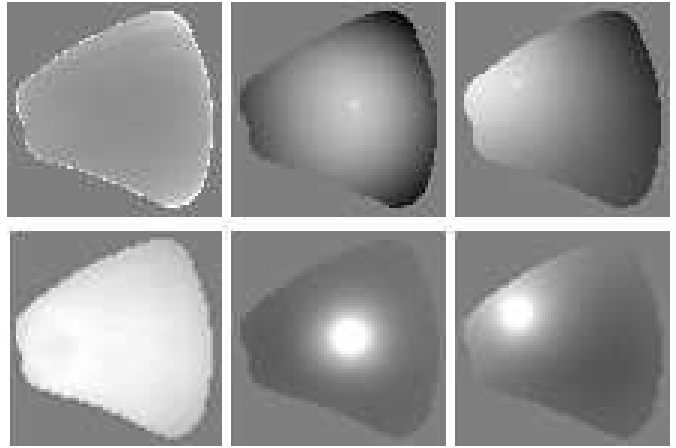}}}}
\caption{
Noise properties of the unsmoothed (top) and 
Wiener filtered (bottom) 
$\gamma$UMi
maps at 43 GHz. The leftmost maps show the 
rms noise in each pixel. The remaining maps show 
the 
correlations with two typical pixels, located at 
the centers of 
the white dots.
}
\label{NoiseCorrFig}
\end{figure}

\section{RESULTS}

\subsection{1/f-noise removal}

\Fig{MethodFig} (left panel) shows the 43 GHz {\AL} map 
obtained with the 
straight averaging method 
(simply averaging the observations of each 
pixel), and illustrates
why the more sophisticated minimum noise method
(middle panel) is needed.
The visible striping along the circular scans 
(see S2000 for details about
the scan pattern) is caused by 
$1/f$-noise, which creates strong correlations 
between map pixels that 
are scanned within a short time of one another.
The middle panel shows that our improved 
mapmaking technique  
has eliminated this striping. This is more 
vividly illustrated in \fig{NoiseCorrFig}
(upper right panels), showing two typical rows of 
the noise
covariance matrix $\NN$. Since row $i$ gives the 
correlations to
$\ith$ pixel, the small round white dots tell us 
that strong 
noise correlations persist only between pixels 
that are close neighbors.
Moreover, these correlations are seen to be quite 
isotropic, with no
signs of elongation along scan directions (the 
only exceptions
are pixels close to edges). 
There is a slight anticorrelation to very distant
pixels caused by the above-mentioned monopole 
removal, 
which forces the correlation map to have a 
vanishing average.
This removal of 1/f-noise also causes a 
substantial reduction in the
{\rms} pixel noise (upper left panel in Figure 
5), which is about a factor of two 
lower than for the straight averaging map.
Both of these desirable properties of the pixel 
noise can be readily understood
by comparing our scan strategy to the four toy 
models described in 
T97c. Since the circular HACME scans cross one 
another at virtually 
every pixel, the geometry is similar to the 
``fence pattern'', which was found 
to produce isotropic noise correlations like 
those in \fig{NoiseCorrFig}.
A ``fence'' map made with the straight averaging 
method 
would correlate
the noise in a pixel with neighbors 
along the two perpendicular scan directions, 
which form a  
``+''-shaped region around the pixel.
With the minimum-noise method, on the other hand, 
a pixel is linked to
{\it all} its neighbors via a sequence of 
intersecting scans in a zig-zag pattern. This is 
accomplished automatically in the matrix 
inversion step,
and has the effect of isotropizing correlations 
(thereby 
reducing long-range correlations) and utilizing
the available information better (reducing the 
resulting 
pixel noise).

\subsection{Wiener-filtered maps}

The {\rms} pixel noise 
$\sigma_i\equiv\NN_{ii}^{1/2}$
in the unsmoothed 43 GHz {\AL} map 
(middle panel of \fig{MethodFig}) is about $400\>\mK$ 
per $(0.3^\circ)^2$ pixel
in the central parts and $250\>\mK$ when rebinned 
into
$(0.8^\circ)^2$ pixels, {\ie}, 
substantially larger than any expected CMB 
signal. What the eye perceives 
is therefore mostly noise, so it is desirable to 
increase the signal-to-noise
ratio by smoothing. Moreover, some pixels near 
the edges are extremely noisy
(with an {\rms} around $10\>\milliK$) since they 
were only 
observed a small number of times (upper left 
panel of \fig{NoiseCorrFig}). 
To prevent them from dominating the image,
it is desirable to give them less weight.
The Wiener filtered map (right panel of \fig{MethodFig}) 
automatically remedies both
of these problems. 
The new noise map (lower left panel of \fig{NoiseCorrFig}) 
shows that the noise explosion
near the edges has been eliminated, 
and the new correlation maps (lower right panels 
of \fig{NoiseCorrFig})
show that smoothing has widened the correlated 
region.
To avoid imprinting features on any particular 
angular scale, we have used a flat
fiducial power spectrum $C_\l = 24\pi 
Q^2/5\l(\l+1)$ in \eq{SdefEq}
just as in T97d, normalized to $Q=30\mK$.
\Fig{WienerFig} shows the Wiener-filtered {\GUM} and 
{\AL} maps for all three channels,
on a symmetric linear grey scale from low (black) 
to high (white) 
temperatures.
Since some maps are substantially noisier than 
others and the Wiener
filtering suppresses their temperature scale 
accordingly, 
the grey scale has been separately normalized for 
each map to avoid 
saturation and allow 
direct comparison of spatial morphology. 

\subsection{Verification and testing}

We tested our data-analysis pipeline by 
generating mock 
sky maps that were observed with the actual HACME 
scan pattern.
These simulated TOD sets were then run through the 
pipeline,
recovering the original maps. When adding 
noise to the TOD with the measured HACME power 
spectrum,
the pixel noise in the recovered map was found to 
be fully consistent 
with the noise covariance matrix $\NN$ computed 
by the pipeline.

To test if the reconstructed map was sensitive to 
the pixel size,
the analysis was repeated for a range of pixel 
sizes. 
As expected, this produced virtually identical 
maps, as long as
the pixel size was smaller than the $0.77^\circ$ 
beam size by
at least the Shannon oversampling factor $2.5$.

To test if the Wiener-filtered maps were sensitive 
to the
choice of fiducial band power, they were computed 
for the
band power normalizations $Q=20$, $30$ and 
$40\,\mK$.
The visual difference between the two extreme 
cases
was minimal, corresponding merely to an increase 
in the effective 
smoothing scale by about $10\%$ (for more details 
on
why this is expected, see the discussion in 
T97d). 

\subsection{Constraints on systematics and the 
CMB power spectrum}

The initial version of our analysis assumed that 
the scan-synchronous
offsets were time-independent. The result was a 
marginal detection of
fluctuations in the most sensitive channel (43 
GHz) and upper
limits at 39 and 41 GHz. However, subsequent 
tests showed that 
variations in these offsets at the level of a few 
percent could
masquerade as signal at this level. Since the 
offsets are completely dominated
by the first harmonic (at the scan frequency), we 
repeated the entire
analysis after throwing away all the information 
about this frequency.
This corresponded to replacing the whitening 
filter $D$
discussed in \sec{MethodSec} by a notch filter that vanished 
at this frequency (as well as at 0 Hz).
The Wiener-filtered maps shown in 
Figures~\ref{ZoomFig} 
and~\ref{MethodFig} were made without the notch filter, 
whereas those 
in \fig{WienerFig} used it.
The result of the notch filter was an increase in 
the rms noise
of about 50\% 
for the least noisy pixels in the raw maps, 
as well as a reduction of the signal seen to 
levels that 
were no longer statistically significant.
The pixel noise $\sigma_i$ (from detector noise 
only)
is $385\,\mK$ in the best pixel in the 43 GHz map 
of the
{\GUM} region,
The 41 and 39 GHz channels are noisier by factors 
of 1.2 and 1.9, respectively. 
For comparison, a CDM-like power spectrum 
normalized 
to a band power $Q_{\rm flat}=30,\mK$
would give corresponding pixel fluctuations of 
order 
$80\,\mK$ in the this map including the mean 
removal.
Most of the fluctuations seen in the raw maps are 
therefore noise 
rather than sky signal, and as described below, 
this remains true
for the Wiener-filtered maps as well.
The reason that many features are seen to recur 
between
the channels is that their noise is correlated.
Although there is a slight sensitivity to the 
dipole 
(with sky rotation marginally breaking the 
degeneracy between 
the scan-synchronous offset and the dipole), 
the resulting noise error bars on this mode are 
too large to allow a detection
after our notch filtering.
     
\begin{figure*}[phbt]
\centerline{{\vbox{\epsfxsize=18cm\epsfbox{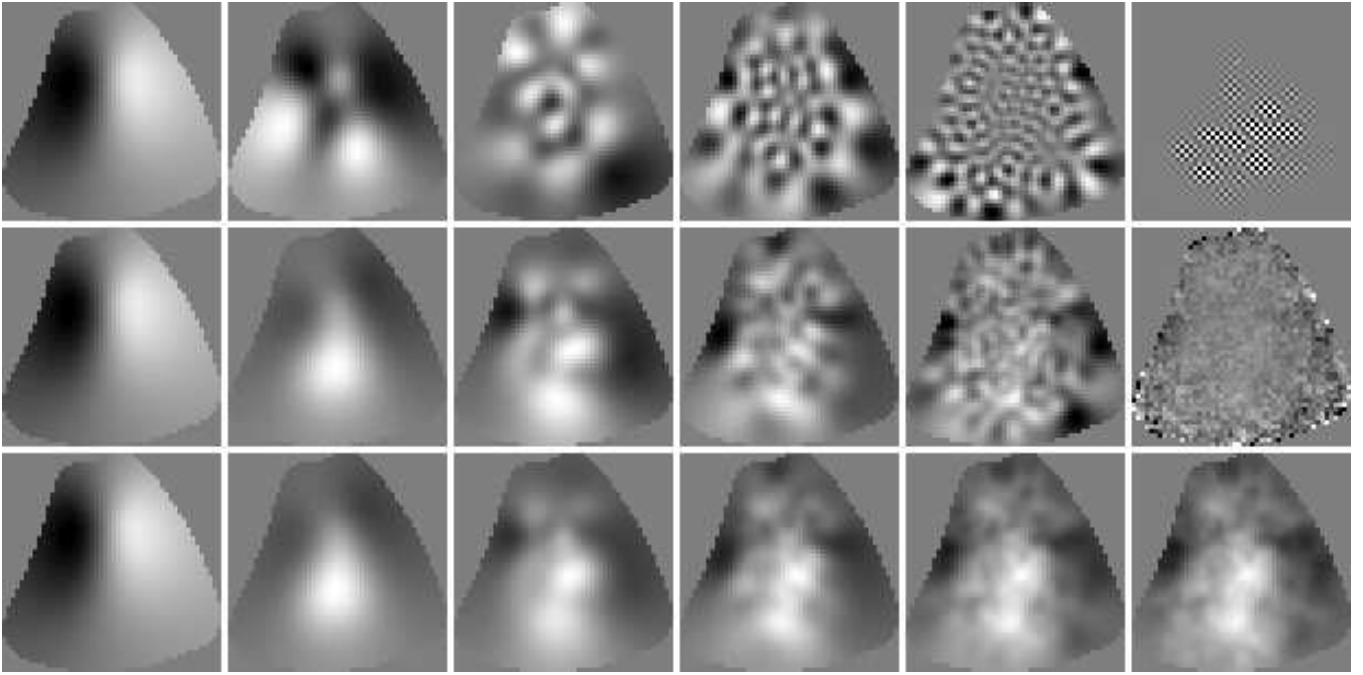}}}}
\caption{
The top row shows a S/N eigenmodes number 1, 3, 
10, 100, 300
and 3082 
for the 43 GHz $\gamma$UMi map,
which are seen to probe successively smaller (and 
noisier) scales.
The second row shows cumulative sums up to these 
mode numbers,
with each mode weighted by its expansion 
coefficient so that the
total sum recovers the original map (right). 
The last row shows the analogous running sums 
with 
the $\ith$ coefficient rescaled by 
$\lambda_i/(\lambda_i+1)$,
so that the total sum recovers the Wiener-
filtered map (right).
}
\label{ModeFig}
\end{figure*}

\begin{figure}[phbt]
\vskip-1cm
\centerline{{\vbox{\epsfxsize=9cm\epsfbox{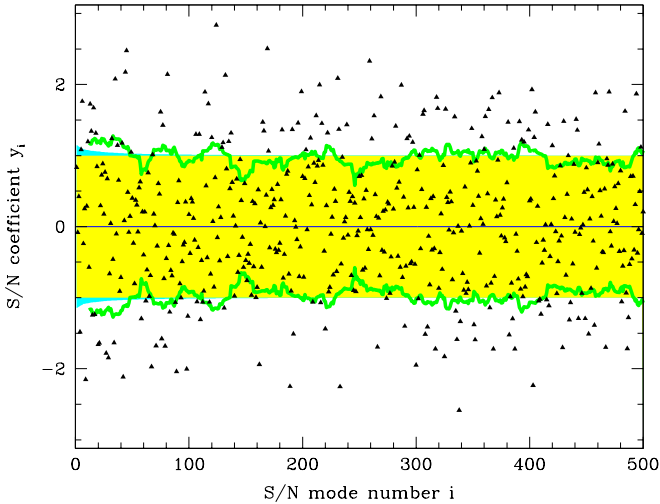}}}}
\vskip-1cm
\caption{
The triangles show the first 500 S/N eigenmode expansion coefficients
$y_i$
for the 43 GHz {\GUM} map.
If there were no systematics or CMB fluctuations in the map,
merely detector noise, 
they would have unit variance, and about $68\%$ of them would would be
expected to lie within the yellow/light grey band.
For a flat $Q_{\rm flat}=30\,\mu$K CMB power spectrum,
the standard deviation would be larger for the best modes, 
as indicated by the 
thin blue/grey band on the left side. The wiggly green/grey 
curve is the rms of the data points
$y_i$, averaged in bands of width 25.
}
\label{SNfig}
\end{figure}

To place quantitative constraints on the 
systematics and sky signal
in the maps, we perform a signal-to-noise 
eigenmode analysis
(Bond 1994; Bunn \& Sugiyama 1995).
We define a new data vector
\beq{zDefEq}
\y\equiv\B^t\x,
\eeq
where $\b$, the columns of the matrix $\B$, are 
the
eigenvectors of the generalized eigenvalue problem
\beq{SNeigenEq}
\S\b = \lambda \NN\b,
\eeq
sorted from highest to lowest eigenvalue $\lambda$
and normalized so that $\b^t\N\b=\I$.
A sample of eigenmodes $\b$ are shown in 
\fig{ModeFig},
and the corresponding expansion coefficients $y_i$
are plotted in \fig{SNfig}.
If there are no systematic errors or residual 
offset
problems, detector noise should 
contribute uncorrelated fluctuations of unit 
variance 
to each mode. All systematic problems we have 
considered would
add rather than subtract map power.
\Fig{SNfig} indeed shows no evidence
of outliers or other problems with our noise 
model.

The covariance of the expansion coefficients is
$\expec{y_i y_j}=\delta_{ij}(1+\lambda_i)$,
so cosmic signal at the assumed level would boost 
the
variance somewhat. The very best 
43 GHz {\GUM} mode $(i=1)$ has $\lambda_i\approx 
0.32$,
corresponding to an expected {\rms} 
of $(1+\lambda_i)^{1/2}\approx 1.15$ for $y_i$.
\Fig{SNfig} is seen to be quite consistent with 
this.
Thus no single mode has signal-to-noise exceeding 
unity.

\begin{figure}[phbt]
\vskip-1cm
\centerline{{\vbox{\epsfxsize=9cm\epsfbox{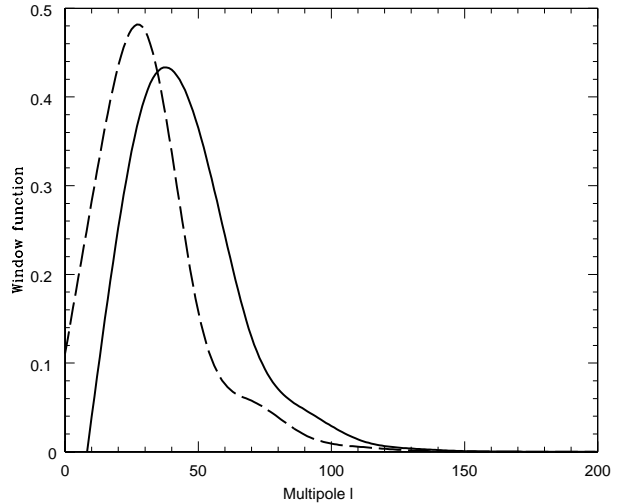}}}}
\vskip-1cm
\caption{
The window functions are shown for the analysis 
with (solid) and without (dashed)
the notch filter. The latter is seen to destroy 
sensitivity on scales 
$\ell\simlt 20$.
}
\label{WindowFig}
\end{figure}

To place cosmological constraints, we need to combine the information
from all modes. We do this using the decorrelated quadratic estimator
method described in Tegmark (1997b). We find that no channel gives 
a statistically significant detection. The sharpest upper limit comes from
the 43 GHz {\GUM} data, giving 
$\delta T\equiv [\l(\l+1)C_\l/2\pi]^{1/2}<77\,\mK$ at 95\% confidence
with the window function shown in \fig{WindowFig}. 
This corresponds to the effective angular scale $\l=38^{+25}_{-20}$. 

What was the price paid for using the notch filter? \Fig{WindowFig}
shows that this destroys information on scales $\ell\simlt 20$,
so the 50\% increase in rms pixel noise that we found above
is caused largely by increased detector noise fluctuations on 
these large angular scales, comparable to the size of the 
basic $10^\circ$ scanning circle.

\section{CONCLUSIONS}

We have presented CMB maps from the HACME balloon 
experiment,
which was devised as a pathfinder
for the long-duration balloon experiments BEAST 
and ACE.
Our analysis has clearly demonstrated that 
neither substantial 
$1/f$-noise nor slightly variable scan-synchronous 
offsets
are ``show-stoppers'', but 
can be efficiently dealt with using the matrix-based 
minimum variance notch filter technique.
However, to provide interesting cosmological 
constraints, the raw sensitivity needs to be 
substantially
improved over the HACME pathfinder specifications.
Comparing this with the straight averaging method
(averaging the observations of each pixel), 
we found that our analysis technique improved the 
results in two ways:
\begin{enumerate}
\item The resulting {\rms} noise in the maps is 
reduced by
about a factor of two, the equivalent of four 
times
more flight time.
\item Long-range anisotropic noise correlations 
(``striping'') is virtually eliminated from the 
maps.
\end{enumerate}
This bodes well for upcoming spin-chopped total 
power 
experiments such as BEAST and ACE.
BEAST is anticipated to have nearly 3 times 
higher 
sensitivity in $\mK\>\second^{1/2}$ and  
2-6 times the angular resolution, with the same 
ratio
of $1/f$ knee to spin rate as HACME and 50 times 
more
integration time. Indeed, the notch filter 
approach should be
useful for any ground- or balloon-based CMB 
experiment 
using an interconnected scan-strategy.

\bigskip
Support for this work was provided by
NASA through grants NAG5-6034 and NAGW-1062 and
through a Hubble Fellowship,
HF-01084.01-96A, awarded by the Space Telescope 
Science
Institute, which is operated by AURA, Inc. under 
NASA
contract NAS5-26555, and by the NSF 
Center for Particle Astrophysics.
JWS was supported by GSRP grant NGW-51381.
N.F. was partially supported by CNPq and CAPES.




\end{document}